\documentclass[graybox]{svmult}

\usepackage{mathptmx}       
\usepackage{helvet}         
\usepackage{courier}        
\usepackage{type1cm}        
\usepackage{makeidx}         
\usepackage{graphicx}        
\usepackage{multicol}        
\usepackage[bottom]{footmisc}

\makeindex             

\begin{document}

\title*{Towards precise asteroseismology of solar-like stars}
\titlerunning{TDC modelling of the solar-like star $\beta$ Hydri} 
\author{A.~Grigahc\`ene, M.-A.~Dupret, S.~G.~Sousa, M.~J.~P.~F.~G.~Monteiro, R.~Garrido, R.~Scuflaire and M.~Gabriel}
\authorrunning{A.~Grigahc\`ene et al.}

\institute{
A.~Grigahc\`ene, S.~G.~Sousa and M.~J.~P.~F.~G.~Monteiro \at Centro de Astrof\'{\i}sica da Universidade do Porto, Rua das Estrelas, 4150-762 Porto, Portugal 
\and M.-A.~Dupret, R.~Scuflaire and M.~Gabriel \at Institut d'Astrophysique et de G\'eophysique, Universit\'e de Li\`ege, All\'ee du 6 Ao{\^u}t 17-B 4000 Li\`ege, Belgium 
\and  M.~J.~P.~F.~G.~Monteiro \at Departamento de F\'{\i}sica e Astronomia, Faculdade de Ci\^encias da Universidade do Porto, Porto, Portugal 
\and R.~Garrido \at Instituto de Astrof\'{i}sica de Andaluc\'{\i}a, CSIC, PO Box 3004, 18080 Granada, Spain
}

%
%
\maketitle

\abstract{
Adiabatic modeling of solar-like oscillations cannot exceed a certain level
of precision for fitting individual frequencies. This is known as the problem 
of near-surface effects on the mode physics. We present a theoretical 
study which addresses the problem of frequency precision in non-adiabatic
models using a time-dependent convection treatment.  We find that the 
number of acceptable model solutions is significantly reduced and more 
precise constraints can be imposed on the models.  Results obtained for a 
specific star ($\beta$ Hydri) lead to very good agreement with both global
and local seismic observables. This indicates that the accuracy of model 
fitting to seismic data is greatly improved when a more complete description 
of the interaction between convection and pulsation is taken into account.}

\abstract*{
Adiabatic modeling of solar-like oscillations cannot exceed a certain level
of precision for fitting individual frequencies. This is known as the problem 
of near-surface effects on the mode physics. We present a theoretical 
study which addresses the problem of frequency precision in non-adiabatic
models using a time-dependent convection treatment.  We find that the 
number of acceptable model solutions is significantly reduced and more 
precise constraints can be imposed on the models.  Results obtained for a 
specific star ($\beta$ Hydri) lead to very good agreement with both global
and local seismic observables. This indicates that the accuracy of model 
fitting to seismic data is greatly improved when a more complete description 
of the interaction between convection and pulsation is taken into account.}

\section{Introduction}

Two conditions are necessary for precise probing of the physics of 
stellar interiors with asteroseismology.  One condition concerns the
quality of the observed data.  Space missions have resulted in
data of unprecedented quality and improved data analysis.  The other 
condition is that models should be able to reproduce the complex 
physics at work in these stars, allowing for a precise interpretation of the observations.

The treatment of convection is one of the largest sources of uncertainty in 
stellar astrophysics.  For cool pulsating stars, the 
convective envelope introduces severe complications for modelling the
oscillations. In these stars, the interaction between pulsation
and convection is very important for proper modeling of the oscillations. 
For example, this interaction has been found to be a major source of 
damping in $\delta$~Scuti stars \cite{2005AA...435..927D} and in driving of
oscillations in $\gamma$~Doradus stars \cite{2004AA...414L..17D}. 

For stars with solar-like pulsations, it turns out that the transition 
region where the thermal relaxation time is of the same order as the 
pulsation period is located inside the superficial convective zone. 
Since these near-surface layers have not been properly modeled, they 
give rise to discrepancies between the computed and observed 
frequencies \cite{1988AA...200..213D}.

Gabriel's formalism of time-dependent convection (TDC) \cite{1996BASI...24..233G, 
2005AA...434.1055G}, as implemented in the MAD code \cite{2001AA...366..166D}, 
leads to very important results for models of $\delta$~Sct and $\gamma$~Dor stars 
\cite{2004AA...414L..17D, 2005AA...435..927D} and also for the solar damping rates
\cite{2006ESASP.624E..97D}.

In this work, we present TDC results for models of $\beta$ Hydri, a G2\,IV 
sub-giant showing solar-like oscillations. A recent list of observed pulsation 
frequencies in this star is given in Brand{\~a}o et al. \cite{2011AA...527A..37B}.  
We compare these frequencies with calculations from adiabatic models with and
without the near-surface corrections of Kjeldsen et al. \cite{2008ApJ...683L.175K}.

\section{TDC modeling of $\beta$~Hydri}

\begin{table}[t]
\begin{center}
\caption{Some properties of the best-fitting models.}
{\setlength{\tabcolsep}{4pt}
\begin{tabular}{ l  c  c  c  c  c  c  c  c  c }
\hline
Model        & M (M$_{\odot}$)            & R(R$_{\odot}$)             & Age (Gyr)       & T$_\mathrm{eff} $  $\mathrm{(K)}$   &   Z/X    &   $\alpha_{\mathrm{MLT}}$& $\langle \Delta \nu \rangle$  ($\mu$ Hz)  &   $\chi^2_{\nu}$ & $\chi^2_{\mathcal{P}}$   \\
\hline\noalign{\smallskip}
TDC     & 1.072   &  1.8211  & 8.1957   & 5858   &    0.0194 &   2.300 &  57.43    &    0.14  &  0.26           \\
Adiab.  & 1.070   &  1.8223  & 8.2923   & 5856   &    0.0194 &   2.500 &   57.67    &    0.58  &  0.29      \\
Corr.   & 1.080   &  1.8207  & 7.8221   & 5856   &    0.0194 &   1.900 &   57.63    &   0.40  &  0.31      \\
\hline\noalign{\smallskip}
\end{tabular}
\label{tab}
}
\end{center}
\end{table}

\begin{figure}[t]
\centering
\includegraphics[scale=.5]{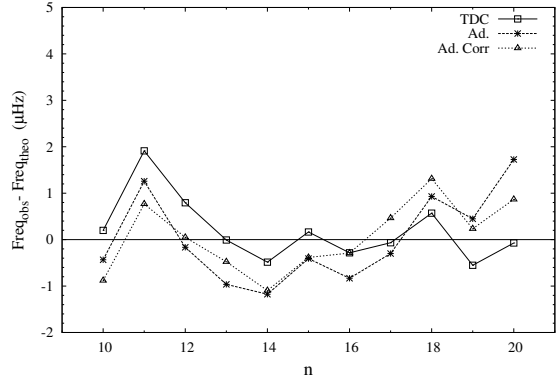}
\caption{The difference between theoretical and observed frequencies for radial modes
using TDC (squares), pure adiabatic (asterisks) and adiabatic with near-surface
corrections (triangles).}
\label{Figfreq}  
\end{figure}

We constructed a grid of TDC models with equilibrium models and non-adiabatic non-radial 
TDC oscillations calculated as described in Dupret et al. \cite{2005AA...435..927D}. 
We then searched for the best-fitting model by maximizing the likelihood:
\begin{eqnarray*}
\mathcal{L} & = & \left( \prod_{j}^{N_{\mathcal{O}}} \frac{1}{\sqrt{2 \pi} \sigma_{j}} 
\right) \times \exp(-\chi^{2}/2),  
\end{eqnarray*}
where $\chi^{2}=\chi^2_{\nu}+\chi^2_{\mathcal{P}}=\frac{1}{N_{\mathcal{O}}} 
\sum^{N_{\mathcal{O}}}_{j=1} \left( \frac{\mathcal{O}^{theo}_{j}-
\mathcal{O}^{obs}_{j}}{\sigma^{\mathcal{O}}_{j}} \right)^{2}$, describes the
match between the ${\mathrm N_{\mathcal{O}}}$ parameters of the theoretical model 
$\mathcal{O}^{theo}$ and the observed values $\mathcal{O}^{obs}$.  These
parameters are the effective temperature, ${\mathrm T_{\rm eff}}$, the mass, 
${\mathrm M}$, the gravity, $\mathrm{log~g}$, the radius, $\mathrm{R}$, and the 
chemical composition, $\mathrm{(X, Z)}$ as well as the frequencies of the radial
modes. The quantities $\chi^2_{\nu}$ and $\chi^2_{\mathcal{P}}$ give the partial 
values of $\chi^{2}$ associated with the frequencies and the fundamental parameters 
respectively.

\begin{figure}[t]
\label{figProb}      
\centering
\includegraphics[scale=.5]{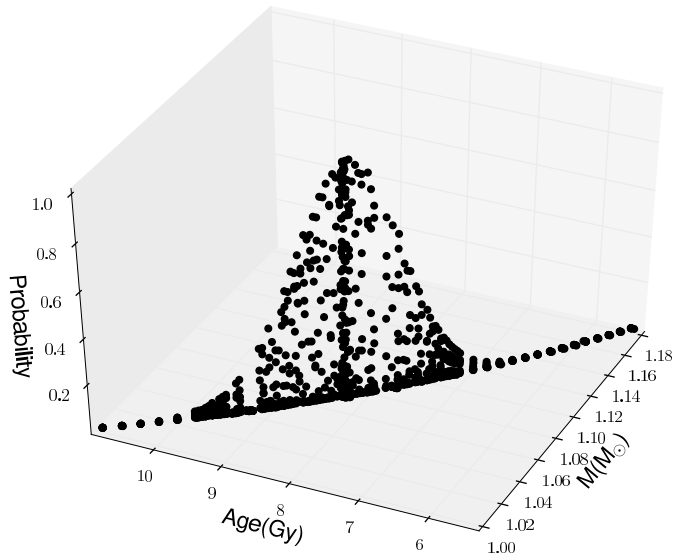}
\caption{The relative probability distribution as a function of mass, $M$, and age.}  
\end{figure}

The difference between the theoretical and observed frequencies is shown in 
Fig.~\ref{Figfreq} and the resulting values of $\chi^2_{\nu}$ listed in column 9 
of Table~\ref{tab}. We note that the adiabatic models give the poorest
agreement with observations. The TDC frequencies agree best with
observations as shown by $\chi^2_{\nu}$. The fact that $\chi^2_{\nu}$
for models with near-surface corrections is much larger than for TDC
models shows that near-surface corrections only apply to stars very similar to the 
Sun.  This is, of course, to be expected since the Sun is used to obtain these 
corrections.

In Fig.~\ref{figProb} we give the probability distribution as a function of mass
and  age. In Table~\ref{tab}, columns 2--7 give the derived global parameters.  
Column 10 lists values of $\chi^2_{\mathcal{O}}$.  The adiabatic models give the
largest radii and the lowest masses. TDC models and models with near-surface
corrections give different masses.  Models with near-surface corrections give
the  lowest ages because they use the smallest values of the mixing-length 
parameter, $\alpha_{\mathrm{MLT}}$. The pure adiabatic models give the highest 
ages for the same reason: they use the highest values of $\alpha_{\mathrm{MLT}}$. 
Column 8 in Table~\ref{tab} gives the average large separations, $\langle \Delta 
\nu \rangle$.  Note that these are all very similar and differ by less than 
1~$\mu$Hz from the observed value of 57.48~$\mu$Hz.  The large separation 
obtained using TDC agrees best with the observed value.

On the other hand the mixing length, $\alpha_{\mathrm{MLT}}$, which is a
free parameter, differs considerably for the different models and is higher
than the calibrated value for the Sun ($\alpha_{\mathrm{MLT}} \sim$ 1.8).  
This agrees with the expected variability of $\alpha_{\mathrm{MLT}}$ across 
the HR diagram. 

\section{Conclusions}
The main results of this work are as follows.  (i) TDC leads to a
significant improvement in the agreement between calculated and observed
frequencies  of solar-like oscillations as compared to the pure adiabatic 
frequencies. (ii) A physically more robust modeling of the mode physics 
near the surface provides better constraints on the global and free
stellar parameters. (iii) Near-surface corrections are probably only
valid for stars most closely resembling the Sun.

\end{document}